\documentstyle[12pt,epsfig,amsfonts]{article}
\def \sech{\mathop{\rm sech}\nolimits}

\begin{document}
\title{The Kink variety in systems of two coupled scalar fields
in two space-time dimensions}
\author{A. Alonso Izquierdo$^{(a)}$,
M.A. Gonz\'alez Le\'on$^{(a)}$ \\ and J. Mateos Guilarte$^{(b)}$
\\ {\normalsize {\it $^{(a)}$ Departamento de Matem\'atica
Aplicada}, {\it Universidad de Salamanca, SPAIN}}\\{\normalsize
{\it $^{(b)}$ Departamento de F\'{\i}sica}, {\it Universidad de
Salamanca, SPAIN}}}

\date{}
\maketitle
\begin{abstract}
In this paper we describe the moduli space of kinks in a class of
systems of two coupled real scalar fields in (1+1) Minkowskian
space-time. The main feature of the class is the spontaneous
breaking of a discrete symmetry of (real) Ginzburg-Landau type
that guarantees the existence of kink topological defects.
\end{abstract}

\section{Introduction}
Research into the mathematical properties and physical meaning of
topological defects in relativistic field theory has increased
sharply since the mid seventies of the twentieth century. There
has also been a parallel development in (non-relativistic)
condensed matter physics. Extended states and phase transitions -
e.g. type II superconductivity- are related to the appearance of
such exotic phenomena. Domain wall defects in the real world can
be thought of as solitary waves propagating in a
$(1+1)$-dimensional universe that self-repeat in the remaining two
dimensions. Thus, investigations on kink nature and behaviour in
$\lambda(\phi^4)_2$ or sine-Gordon models inform us about the
properties of the simplest type of topological defect. Realistic
theories, however, involve more than one scalar field and the
study of (1+1)-dimensional $N$-scalar fields models in this
respect is not only worthwhile but almost mandatory. Examples of
theories with $N>1$, where one might be interested in looking at
topological defects, include the linear sigma model, the
Ginzburg-Landau theory of phase transitions, the supersymmetric
Wess-Zumino model, SUSY QCD, etcetera.

Kinks are time-independent finite-energy solutions of the field
equations that have been thoroughly investigated in the $N=1$
case, see e.g. \cite{Rj}. Much less is known about the kink
variety in systems with two or more scalar fields (the reason for
this is also clearly explained in \cite{Rj}). To the best of our
knowledge, however, there are exceptions:
\begin{itemize}
\item A deformation of the linear $O(2)$-sigma model, christened in
the literature as the ${\rm MSTB}$ model, exhibits a rich variety
of kinks. The characteristics of any of these kink defects as well
as the structure of the variety as a whole have been elucidated in
a long series of papers, see References [2-12]. The moduli space
of kinks in an analogous deformation of the linear $O(3)$-sigma
model has also been fully described in \cite{Aai1}.

\item The search for kinks is tantamount to the solving of a
mechanical problem, which is seldom solvable if $N\geq 2$. In
\cite{Aai2} we described the kinks of two $N=2$ field-theoretical
models associated with completely integrable mechanical systems;
i.e., the same idea that works in the ${\rm MSTB}$ model and its
$N=3$ generalization.

\item  In \cite{Aai3}, the kinks of the Wess-Zumino model are shown
to be given by certain real algebraic curves in the complex plane.

\end{itemize}

Another favorable situation occurs when the field-theoretical
model is the bosonic sector of a supersymmetric system. This is
the case of the Wess-Zumino system  and also happens in a $N=2$
model proposed in \cite{B5}, which has been discussed and applied
to describe several interesting physical contexts in the series of
papers [16-25]. Throughout their work, Bazeia et al. identify only
two kinds of kinks: a topological one, with only the first
component non-null, usually termed as the ${\rm TK}1$ kink, and a
second topological kink that has both components non-null and is
called the ${\rm TK}2$ kink. In contrast with the MSTB model,
where the ${\rm TK}1$ kinks are unstable, \cite{J0}-\cite{J1}, and
decay to the ${\rm TK}2$ kinks, \cite{J2}, in the system of Bazeia
et al. there is an interesting phenomenon of kink degeneracy: the
${\rm TK}1$ and ${\rm TK}2$ kinks have the same classical energy.

The main result to be shown in this paper is that the kink
degeneracy is a continuous one rather than the discrete degeneracy
implicit in [16-25]. We shall find a continuous family of kink
solutions to the classical field equations, all of them
degenerated in energy with the ${\rm TK}1$ and ${\rm TK}2$ kinks.
The existence of this variety of kinks is possible because of the
spontaneous breaking of a discrete internal symmetry group. The
quotient of the kink variety by the symmetry group is the kink
moduli space, a structure parallel to the moduli spaces of gauge
theoretical topological defects as vortices, \cite{wein1}, or
magnetic monopoles, \cite{wein2}.

Identification of the kink variety is achieved through the
solution of first-order, rather than second-order, field
equations. In (1+1)-dimensional scalar field theories, first-order
equations are available if, modulo a global sign, a superpotential
is found. Note that the search for a superpotential  is highly
non-trivial if $N\geq 2$. Bazeia et al., however, proposed a
continuously differentiable superpotential in their model, which
in turn guarantees the stability of any finite energy solution of
the associated first-order system of equations through the
classical Bogomol\'ny-Prasad-Sommerfield argument, \cite{BPS}.

The existence of the superpotential tells us that we can
understand the system as the bosonic sector of an ${\cal N}=1$
(1+1)-dimensional supersymmetric field theory, in which the kinks
play a significant r$\hat{\rm o}$le as BPS states. We shall
analyze the supersymmetric extension of this model in a future
work, but we observe that the dimension of the kink moduli space
in this system is such that the index introduced in \cite{Losev}
is zero, showing that the soliton supermultiplets are long or
reducible.

All the foregoing statements are valid for any value of the single
classically relevant coupling constant in the model. In this paper
we shall show another new result: for certain values of the
coupling constant there exists a second superpotential.
Accordingly, a second system of first-order equations is available
that also admits kink solutions, although the old and new solitons
belong to different topological sectors of the configuration
space. For the critical values where the second superpotential is
found, there are two non-equivalent supersymmetric extensions of
the same bosonic sector.

For most of the critical values the second superpotential fails to
be continuously differentiable at a finite number of points in the
${\Bbb R}^2$ internal space. In these cases, the second
Bogomol\'ny bound is not a topological quantity; it also depends
on the values of the superpotential at the points where it is not
differentiable. Kink orbits that cross those points are unstable
and are solutions of the first-order equations only in one
interval, not on the whole spatial line. Nevertheless, these kinks
are solutions of the second-order equations.

A final comment: in concordance with the lifting of the kink
translational degeneracy, we expect that the kink internal
degeneracy will be removed in second-order in the loop expansion
of the energy in the quantum theory.

The paper is organized as follows. In sections \S 2 and \S 3 we
introduce the BNRT model discussed in \cite{B5} and identify a
one-parametric family of kinks, which includes the TK1 and TK2
kinks, as BPS solutions. In sections \S 4 and \S 5 we investigate
the existence of a second decomposition \'a la Bogomol'nyi. We
find that this is possible for certain values of the coupling
constant, for which we discover a second kink family.

\section{The BNRT model}

In the model introduced in \cite{B5} by Bazeia, Nascimento,
Ribeiro and Toledo, henceforth referred to as the BNRT model, the
scalar field is built from two components
$\chi(y^\mu)=(\chi_1(y^\mu),\chi_2(y^\mu))$ and the dynamics is
governed by the action
\begin{eqnarray}
\bar{S}[\chi]&=&\int d^2 y \left[ \sum_{a=1}^2\partial_\mu \chi_a
\partial^\mu \chi_a- \bar{U}(\chi_1,\chi_2)  \right]
\label{eq:action}\\\bar{U}(\chi_1, \chi_2)&=& \frac{1}{2}
\lambda^2 (\chi_1^2-a^2)^2+ \frac{1}{2} \lambda \mu (\chi_1^2-a^2)
\chi_2^2+\frac{1}{8} \mu^2 \chi_2^4+ \frac{1}{2} \mu^2 \chi_1^2
\chi_2^2 \label{eq:dimpot}
\end{eqnarray}
Here, $\lambda$ and $\mu$ are coupling constants with dimensions
of inverse length and $a^2$ is a non-dimensional parameter. We use
a natural system of units, $\hbar=c=1$. The energy functional is
\begin{equation}
\bar{{\cal E}}[\chi]=\int dy \left[ \frac{1}{2} \left(\frac{d
\chi_1}{d y}\right)^2 + \frac{1}{2} \left(\frac{d \chi_2}{d
y}\right)^2 + \bar{U}(\chi_1,\chi_2) \right] \label{eq:dimener}
\end{equation}
where $\chi(y)=(\chi_1(y),\chi_2(y)) \in {\cal C}=\{ {\rm
Maps}({\Bbb R},{\Bbb R}^2)\, /\, \bar{\cal E}[\chi(y)]< \infty$
\}. Introducing non-dimensional fields, variables and parameters,
$\chi_b = 2 a \phi_b$, $y=\frac{2\sqrt{2}}{a \lambda} x$, and
$\sigma=\frac{\mu}{\lambda}$, we obtain expressions that are
simpler to handle. $\bar{{\cal E}}[\chi_1,\chi_2]=\sqrt{2} a^3
\lambda {\cal E}[\phi_1,\phi_2]$ and the non-dimensional energy
functional $-$which depends on the single classically relevant
coupling constant $\sigma$$-$ is:
\begin{equation}
{\cal E}[\phi]=\int dx \left[ \frac{1}{2} \left(\frac{d \phi_1}{d
x}\right)^2 + \frac{1}{2} \left(\frac{d \phi_2}{d x}\right)^2 +
\left( 4 \phi_1^2+ 2 \sigma \phi_2^2-1 \right)^2 + 16 \sigma^2
\phi_1^2 \phi_2^2 \right] \label{eq:adimener}
\end{equation}
The Euler-Lagrange equations read:
\begin{equation}
\frac{d^2 \phi_1}{d x^2} = 16 \phi_1 \left(4 \phi_1^2 + 2 \sigma
(1+\sigma) \phi_2^2-1\right) \hspace{1.2cm} \frac{d^2 \phi_2}{d
x^2} = 8 \sigma \phi_2 \left( 4(\sigma+1) \phi_1^2 + 2 \sigma
\phi_2^2-1\right) \label{eq:ecuseg}
\end{equation}
Besides the spatial parity and translational symmetries, there is
a global or internal symmetry in this model: the reflection
discrete group ${\Bbb G}={\Bbb Z}_2 \times {\Bbb Z}_2$ generated
by the transformations $\pi_1:(\phi_1,\phi_2) \rightarrow
(-\phi_1,\phi_2)$ and $\pi_2:(\phi_1,\phi_2) \rightarrow
(\phi_1,-\phi_2)$ is also a symmetry subgroup of the system.

We shall focus our attention on the $\sigma>0$ regime, where the
vacuum manifold is:
\[
{\cal M}=\left\{ A_1=(\textstyle\frac{1}{2},0); \,
A_2=(-\textstyle\frac{1}{2},0); \, B_1=(0,
{\textstyle\frac{1}{\sqrt{2 \sigma}}}\,); \, B_2= (0,-
{\textstyle\frac{1}{\sqrt{2 \sigma}}} \,) \right\}
\]
The action of ${\Bbb G}$ on ${\cal M}$ is summarized as follows:
$\pi_1(A_1)=A_2$, $\pi_2(B_1)=B_2$. Therefore, ${\cal M}$ can be
seen as the union of two disjoint vacuum orbits: ${\cal M}=A\sqcup
B$, $A=\{ A_1,A_2\}$, $B=\{ B_1,B_2\}$. The vacuum moduli space
$\bar{\cal M}=\frac{{\cal M}}{{\Bbb G}}$ is a set of two elements,
$\bar{\cal M}= {\bf A}\sqcup {\bf B}$, where ${\bf
A}=\frac{A}{{\Bbb Z}_2\times\{e\}}$, and ${\bf
B}=\frac{B}{\{e\}\times{\Bbb Z}_2}$ . The ${\Bbb G}={\Bbb Z}_2
\times {\Bbb Z}_2$ symmetry of the action (\ref{eq:action}) is
spontaneously broken to the $\{e\} \times {\Bbb Z}_2$ subgroup on
the elements in the $A$ orbit and to the ${\Bbb Z}_2\times \{e\}$
subgroup on the vacua of the $B$ orbit.

Because of the degeneracy and the discreteness of the vacuum
manifold ${\cal M}$, the configuration space is the union of
sixteen topologically disconnected sectors. Keeping in mind the
symmetries of the model, we identify the non-trivial topological
sectors as the {\it AA} topological sector (formed by
configurations of ${\cal C}$ that join the $A_1$ and $A_2$ vacua);
the {\it BB} topological sector (configurations that connect the
$B_1$ and $B_2$ vacua), and the ${\it AB}$ sector (formed by
configurations joining one vacuum in the $A$ orbit with another
vacuum in the $B$ orbit).

\noindent We use the trial orbit method \cite{Rj} to show the
previously known kink solutions to the equations
(\ref{eq:ecuseg}).

\vspace{0.3cm}

\noindent {\bf 1.} The ${\bf TK1}^{AA}$ kink

\vspace{0.2cm}

First, we try the curve
\[
\gamma_{\rm TK1^{AA}}=\left \{\phi_2=0 \hspace{0.3cm},
\hspace{0.3cm} -\textstyle\frac{1}{2}\leq \phi_1\leq
\textstyle\frac{1}{2}\right \}
\]
This condition is compatible with equations (\ref{eq:ecuseg}) and
we find
\[
\phi_1^{{\rm TK1}^{AA}}(x)= \pm \frac{1}{2} \, \tanh 2
\sqrt{2}(x+a) \hspace{1cm} \phi_2^{{\rm TK1}^{AA}}(x)=0
\]
as the one-component topological kinks in the {\it AA}.

\vspace{0.2cm}

\noindent {\bf 2.} The {\bf TK2$^{AA}$} kink:

\vspace{0.2cm}

Second, we try the elliptic orbit
\begin{equation}
\gamma_{{\rm TK}2^{AA}}=\left
\{\phi_1^2+\frac{\sigma}{2(1-\sigma)} \, \phi_2^2 =\frac{1}{4}
\hspace{0.3cm}, \hspace{0.3cm} -\frac{1}{2}\leq
\phi_1\leq\frac{1}{2}\right \} \label{eq:trabaze}
\end{equation}
in (\ref{eq:ecuseg}) and find in the $AA$ topological sector the
two-component topological kinks:
\begin{equation}
\phi_1^{{\rm TK}2^{AA}}(x)= \pm \frac{1}{2} \tanh 2\sqrt{2} \sigma
(x+a) \hspace{1cm} \phi_2^{{\rm TK}2^{AA}}(x)=\pm \,
\sqrt{\textstyle\frac{1-\sigma}{2\sigma}} \sech 2 \sqrt{2} \sigma
(x+a) \label{eq:tk2},
\end{equation}
henceforth referred to as TK2$^{AA}$ kinks.

Note that the orbit (\ref{eq:trabaze}) gives kink curves only in
the $\sigma\in (0,1)$ range because if $\sigma\geq 1$ it becomes a
hyperbole that does not connect the vacua.  Moreover,
(\ref{eq:tk2}) describes four different kinks according to the
choices of the signs and one can obtain one from another by using
the spatial parity and internal reflection symmetries.

The existence of one-component topological kinks -unnoticed in the
literature about the model- in the BB topological sectors is
obvious.

\vspace{0.2cm}

\noindent  {\bf 3.} The {\bf TK1}$^{BB}$ kink:

\vspace{0.2cm}

Third, we try the orbit
\[
\gamma_{{\rm TK1}^{BB}}=\left \{\phi_1=0,
-\frac{1}{\sqrt{2\sigma}}\leq
\phi_2\leq\frac{1}{\sqrt{2\sigma}}\right \}
\]
in the second-order field equations (\ref{eq:ecuseg}). We
immediately find that the finite energy solutions
\[
\phi_1^{{\rm TK1}^{BB}}(x)=0 \hspace{1cm} \phi_2^{{\rm
TK1}^{BB}}(x)=\pm \, \frac{1}{\sqrt{2\sigma}} \, \tanh 2
\sqrt{\sigma} \, (x+a)
\]
are the kinks that connect the $B_1$ and $B_2$ vacua.

\section{The moduli space of kinks in the {\it AA} topological sector}

In [16-25] the authors propose a superpotential for the model:
\begin{equation}
U(\phi_1,\phi_2)=\frac{1}{2} \left(\frac{\partial W}{\partial
\phi_1}\right)^2+\frac{1}{2} \left(\frac{\partial W}{\partial
\phi_2}\right)^2\qquad ; \qquad W(\phi) = 4 \sqrt{2} \left(
\frac{1}{3} \phi_1^3 -\frac{1}{4} \phi_1 + \frac{\sigma}{2} \phi_1
\phi_2^2 \right) \label{eq:superpot}
\end{equation}
\noindent\begin{figure}[htbp] \centerline{\epsfig{file=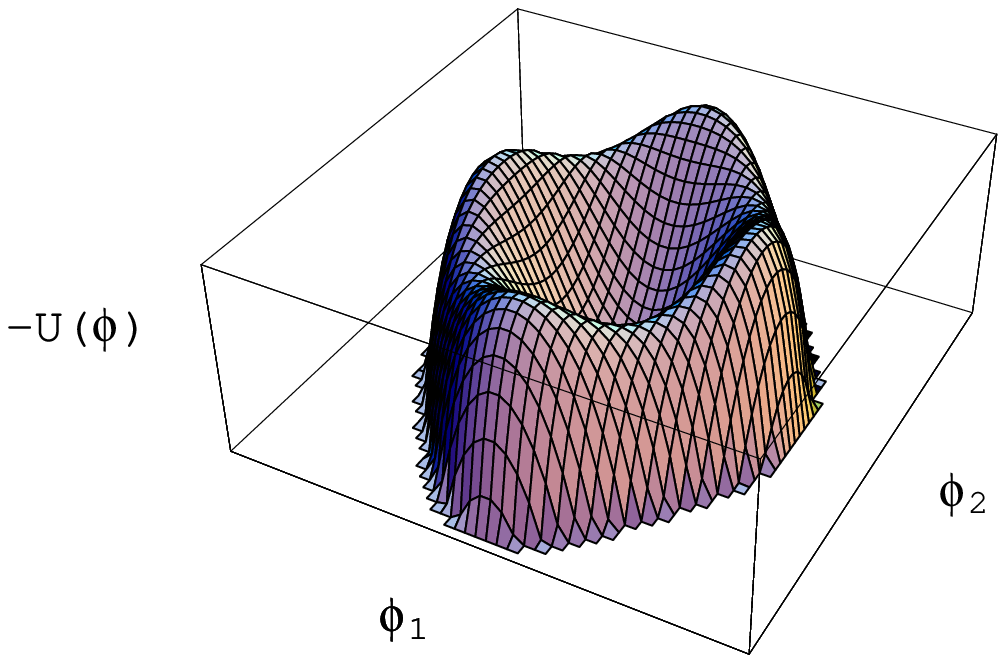,
height=4.5cm} \hspace{1.2cm}
\epsfig{file=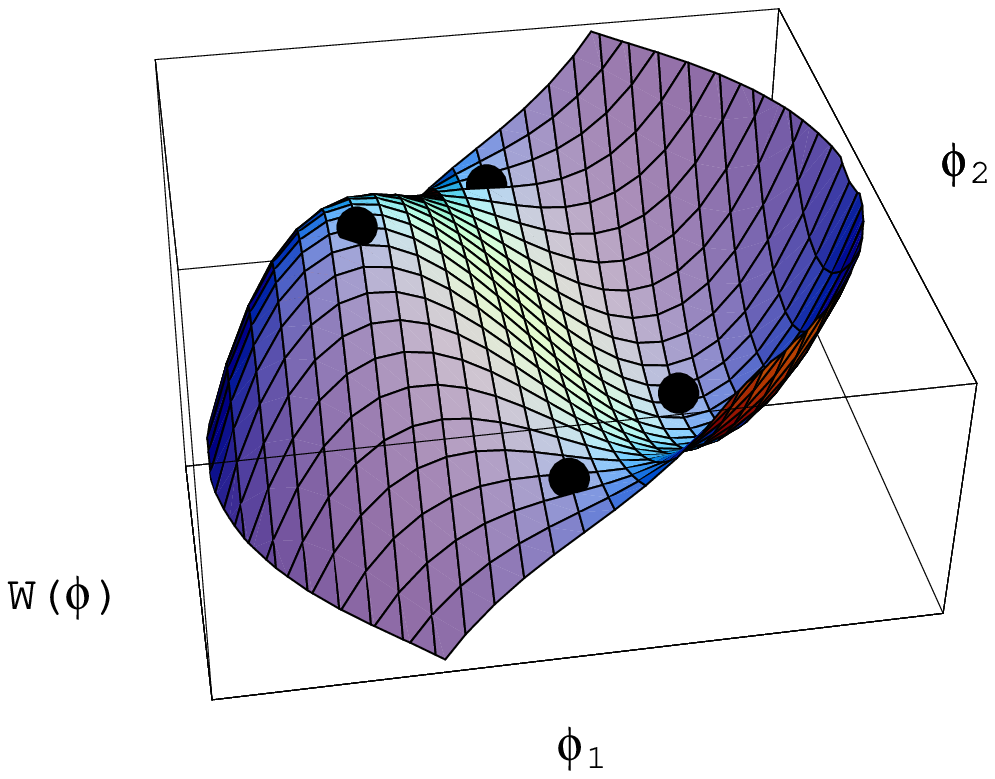,height=4.5cm}} \caption{\small \it The
$U(\phi)$ potential (left) and the superpotential $W(\phi)$
(right)}
\end{figure}

The classical BPS states satisfy the system of first-order
equations
\begin{equation}
\frac{d \phi_1}{d x} =\frac{\partial W}{\partial \phi_1}= \sqrt{2}
(4 \phi_1^2+2 \sigma \phi_2^2-1) \hspace{1cm};\hspace{1cm} \frac{d
\phi_2}{d x} =\frac{\partial W}{\partial \phi_2}=4 \sqrt{2} \sigma
\phi_1 \phi_2 \label{eq:trabaz}
\end{equation}
which are easier to solve than (\ref{eq:ecuseg}). The
superpotential $W(\phi_1,\phi_2)$ is a smooth function of the
fields $\phi_1$ and $\phi_2$ at each point in ${\Bbb R}^2$.
Therefore, according to the Bogomol'nyi arrangement
\[
{\cal E}[\phi]=\int dx \sum_{a=1}^2 \left( \frac{d \phi^a}{d
x}-\frac{\partial W}{\partial \phi^a} \right)^2+\int
\frac{\partial W}{\partial \phi^a} \frac{d \phi^a}{dx}
\]
we have that
\[
{\cal E
}[\phi]=T[\phi]=|W(\phi_1(\infty),\phi_2(\infty))-W(\phi_1(-\infty),\phi_2(-\infty))|
\]
for all solutions of (\ref{eq:trabaz}) and the kink energy only
depends on the topological sector of the solution.

The kink solutions of (\ref{eq:trabaz}) are the flow-lines of
${\rm grad}\,W$ that start and end at elements of ${\cal M}$. It
happens that $A_1$ and $A_2$ are respectively maxima and minima of
$W$ and that there are flow-lines of ${\rm grad}\,W$ starting at
$A_1$ and ending at $A_2$ (or vice-versa). $B_1$ and $B_2$,
however, are saddle points of $W$, see Figure 1. Therefore, there
are no flow-lines of ${\rm grad}\,W$ between $B_1$ and $B_2$ (or
vice-versa). Nevertheless, flow-lines of ${\rm grad}\,W$ between
one point in the $A$ orbit and another point in the $B$ orbit (or
vice-versa) are possible. The flow-lines of ${\rm grad}\,W$ thus
provide kinks in the ${\it AA}$ and the ${\it AB}$ sectors with
energies:$E_{{\rm TK}2^{AA}}=\frac{4}{3}a^3\lambda$, $E_{{\rm
TK}2^{AB}}=\frac{2}{3}a^3\lambda$.

To obtain the most general solution to the first-order system
(\ref{eq:trabaz}), we first integrate the first-order ODE
\begin{equation}
\frac{d \phi_1}{d \phi_2}= \frac{4 \phi_1^2+2 \sigma \phi_2^2
-1}{4 \sigma \phi_1 \phi_2} \label{eq:orbflu}
\end{equation}
which admits the integrating factor
$|\phi_2|^{-\frac{2}{\sigma}}\phi_2^{-1}$, if $\sigma\neq 1$ and
$\sigma\neq 0$, thereby allowing us to find all the flow-lines as
the family of curves
\begin{equation}
\phi_1^2 + \frac{\sigma}{2 (1-\sigma)} \phi_2^2 = \frac{1}{4}+
\frac{c}{2 \sigma} |\phi_2|^{\frac{2}{\sigma}} \label{eq:traba1}
\end{equation}
parametrized by the real integration constant $c$. There is a
critical value
\[
c^S=\frac{1}{4} \frac{\sigma}{1-\sigma} \left(
2\sigma\right)^{\frac{\sigma+1}{\sigma}}
\]
and the behaviour of a particular curve in the (\ref{eq:traba1})
family is described in the following items:
\begin{itemize}
\item For $c \in (-\infty,c^S)$, formula (\ref{eq:traba1}) describes
closed curves in the internal space ${\Bbb R}^2$ that connect the
vacua $A_1$ and $A_2$, see Figure 2. Thus, they provide a kink
family in the topological sector {\it AA}. Henceforth, we refer to
these kinks as {\bf TK2}$^{AA}(c)$. A fixed value of $c$
determines four members in the kink variety related amongst one
another by spatial parity and internal reflections. The kink
moduli space is defined as the quotient of the kink variety by the
action of the symmetry group:
\[
{\cal M}_{\rm K}=\frac{{\cal V}_{\rm K}}{{\Bbb P}\times{\Bbb
G}}=(-\infty , c^S ),
\]
the real open half-line parametrized by $c$. One sees that
\[
{\bf TK2}^{AA} \equiv {\bf TK2}^{AA}(0) \hspace{2cm} {\bf TK1}
\equiv \lim_{c \rightarrow -\infty} {\bf TK2}^{AA} (c)
\]
i.e. the ${\rm TK}2^{AA}$ kink is the $c=0$ member of the family
(if $\sigma<1$) and the ${\rm TK}1^{AA}$ kink is not strictly
included although it does appear at the boundary of ${\cal M}_{\rm
K}$.

\item In the range $c\in (c^S,\infty)$, equation (\ref{eq:traba1})
describes open curves and no vacua are connected. These ${\rm
grad}\,W$ flow-lines are infinite energy solutions that do not
belong to the configuration space ${\cal C}$, see Figure 2.

\item At the other point of the boundary of ${\cal M}_{\rm K}$, $c=c^S$,
we find the ${\rm TK}2^{AB}$ kinks, which are the separatrices
between bounded and unbounded motion and the envelop of all kink
orbits in the {\it AA} topological sector, see Figure 2.
\end{itemize}

We briefly discuss the $\sigma=1$ case. The $\sigma=0$ case is not
interesting because the $\phi_2$ dependence disappears in the
potential: it is a \lq\lq direct sum" of an $N=1$ $\phi^4$ model
and an $N=1$ free model. Integration of $(\ref{eq:orbflu})$ when
$\sigma=1$ gives
\begin{equation}
\phi_1^2-\phi_2^2 \left( \frac{c}{2}+\log |
\phi_2|\right)=\frac{1}{4} \label{eq:traba2}
\end{equation}
where the kink trajectories now appear in the $c\in (-\infty,
c^S]$ range, with $c^S=-1+\ln 2$. The description of the kink
orbits is analogous to the description for $\sigma\neq 1$ above.
\begin{figure}[htbp]
\centerline{\epsfig{file=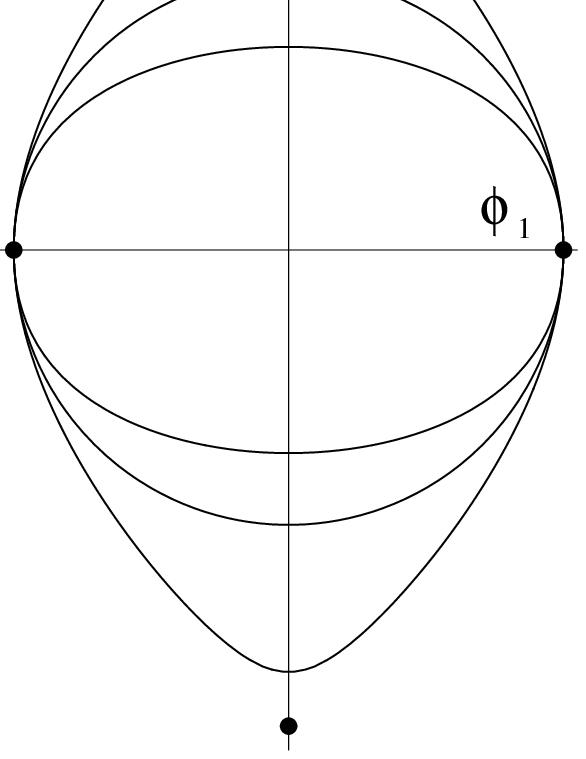, height=4.cm} \hspace{1.5cm}
\epsfig{file=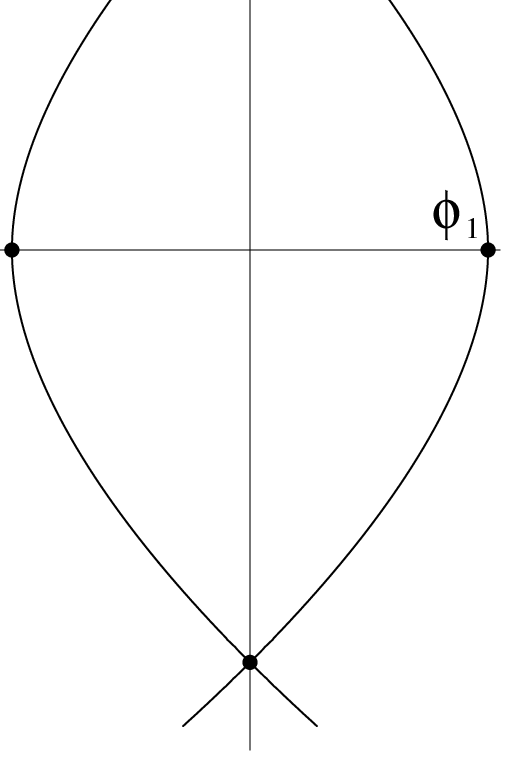,height=4.cm}
\hspace{1.5cm}\epsfig{file=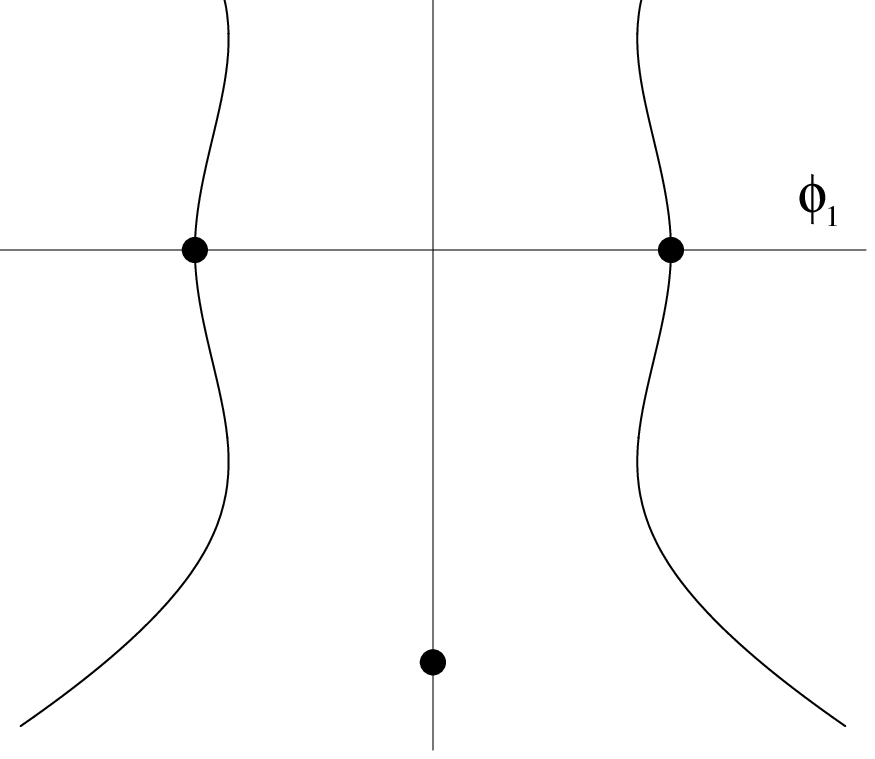, height=4.cm}}
\caption{\small \it Flow-lines given by (\ref{eq:traba1}): for $c
\in (-\infty,c^S)$ (left), $c=c^S$ (middle), and $c \in (c^S
,\infty)$ (right).}
\end{figure}

A second step remains: the explicit dependence of the kinks with
respect to the space coordinate can be obtained if we plug
(\ref{eq:traba1}) into the second equation in (\ref{eq:trabaz}),
\begin{equation}
h[\phi_2]=\int \frac{d \phi_2}{\phi_2 \sqrt{\frac{1}{4}+\frac{c}{2
\sigma} |\phi_2|^{\frac{2}{\sigma}}-\frac{\sigma}{2(1-\sigma)}
\phi_2^2}} = \int 4 \sqrt{2} \sigma dx  \label{eq:integral} .
\end{equation}
The kink solutions are
\[
\phi_1^{\rm K}(x,c)= \pm \sqrt{\frac{1}{4}+\frac{c}{2 \sigma}
|h^{-1}(4 \sqrt{2} \sigma
x)|^{\frac{2}{\sigma}}-\frac{\sigma}{2(1-\sigma)} [h^{-1}(4
\sqrt{2} \sigma x)]^2}\hspace{1cm} \phi_2^{\rm K}(x,c)=h^{-1}(4
\sqrt{2} \sigma x)
\]
In general, we cannot obtain the explicit dependence on $x$ for
the kink solutions because either we cannot integrate
(\ref{eq:integral}) or we cannot identify the inverse of
$h(\phi)$. For certain values of the coupling constant, however,
we can finish the task. We next show the family of ${\rm
TK}2^{AA}$ kinks for $\sigma=2$ and $\sigma=\frac{1}{2}$.

\vspace{0.3cm}

$\bullet$ ${\bf\sigma}=2$:

\vspace{0.2cm}

The vacuum points are the vertices of a square: ${\cal
M}_{\sigma=2}=\{A_1=(\frac{1}{2},0),A_2=(-\frac{1}{2},0),
B_1=(0,\frac{1}{2}), B_2=(0,-\frac{1}{2})\}$. The quadratures
(\ref{eq:integral}) can be solved explicitly and $h^{-1}[\phi_2]$
is a known analytical function. Thus,
\[
\phi_1^{{\rm TK2}^{AA}}(x)=\pm \frac{1}{2} \frac{\sinh 4
\sqrt{2}(x+a)}{\cosh 4 \sqrt{2} (x+a)+b} \hspace{1cm} \phi_2^{{\rm
TK2}^{AA}}(x)=\pm \frac{1}{2} \frac{\sqrt{b^2-1}}{\cosh 4 \sqrt{2}
(x+a)+b}
\]
are the kink-form factors. The integration constant $b$ is related
to $c$ as $b=\frac{-c}{\sqrt{c^2-16}}$, and for $b\in (1,\infty)$
we find kinks in the {\it AA} topological sector.

If $c=c^S=-4$, $b=\infty$ we find the kinks in the {\it AB} sector
\[
\phi_1^{{\rm TK2}^{AB}}(x)=\pm \frac{1}{4} (1-\tanh 2 \sqrt{2}
(x+a))
 \hspace{1cm} \phi_2^{{\rm TK2}^{AB}}= \pm
\frac{1}{4}(1+\tanh 2 \sqrt{2} (x+a))
\]
and, replacing $x$ by $-x$, its antikinks. The separatrices are
placed on the edges of the above mentioned square $\phi_2=\pm
\frac{1}{2}\pm \phi_1$. The kink trajectories in the {\it AA}
topological sector form a dense family of curves enveloped by the
kink orbits in the {\it AB} sector. See Figure 3.

\begin{figure}[htbp]
\centerline{\epsfig{file=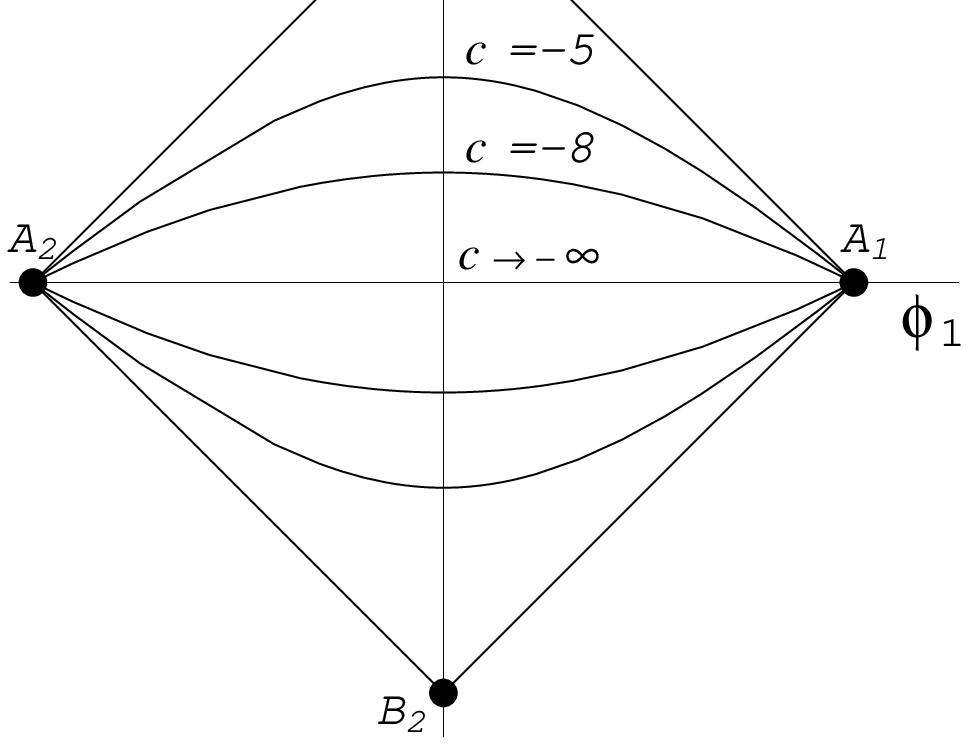, height=4.5cm} \hspace{0.5cm}
\epsfig{file=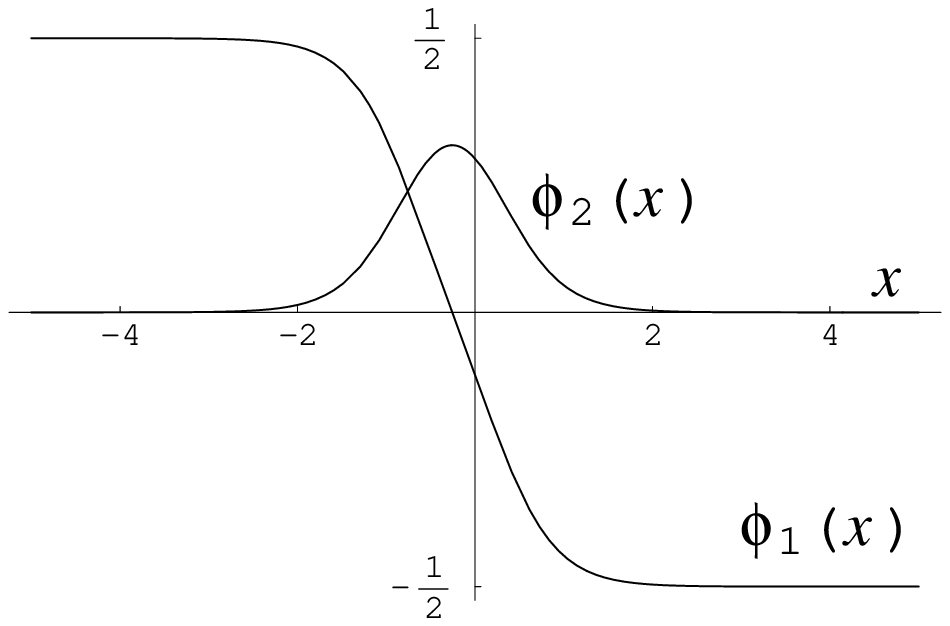,height=3.5cm}
\hspace{0.5cm}\epsfig{file=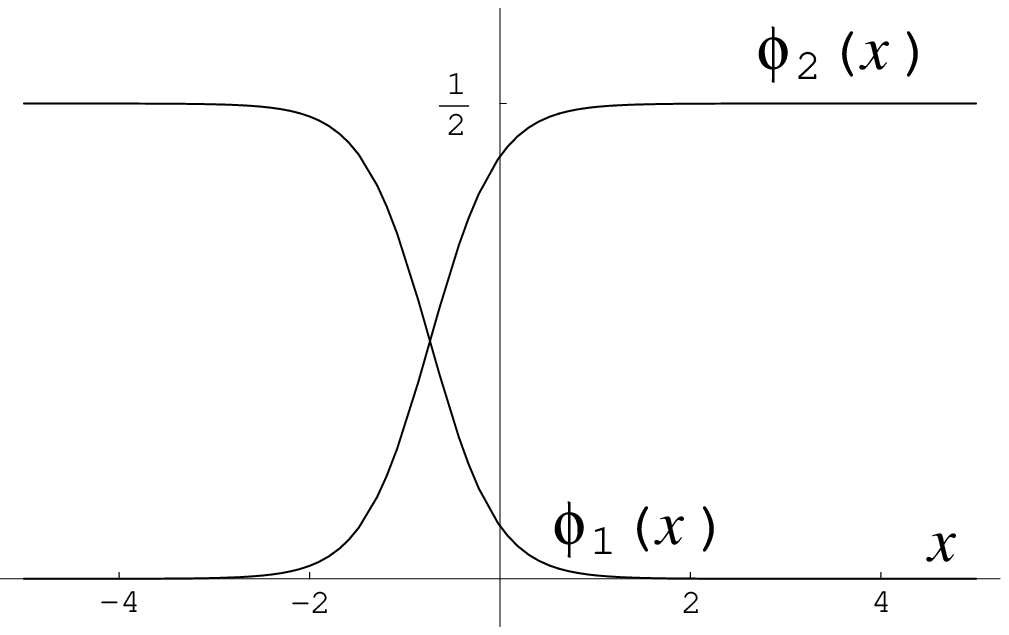, height=3.5cm}}
\caption{\small \it Kink trajectories (left), a kink in the AA
sector (middle) and a kink in the AB sector (right) in the case
$\sigma=2$.}
\end{figure}

A rotation of $45^o$ in ${\Bbb R}^2$,
$\phi_1=\frac{1}{\sqrt{2}}(\psi_1+\psi_2)$ and
$\phi_2=\frac{1}{\sqrt{2}}(\psi_1-\psi_2)$, shows that for this
value of $\sigma$ the system is non-coupled:
$U_{\sigma=2}(\psi_1,\psi_2)=\frac{1}{32}
(\psi_1^2-\frac{1}{8})^2+\frac{1}{2} (\psi_2^2-\frac{1}{8})^2$.

\vspace{0.3cm}

$\bullet$ ${\bf\sigma}=\frac{1}{2}$:

\vspace{0.2cm}

The vacuum manifold is: ${\cal
M}_{\sigma=\frac{1}{2}}=\{A_1=(\frac{1}{2},0),A_2=(-\frac{1}{2},0),
B_1=(0,1), B_2=(0,-1)\}$. By the same procedure as above, we
obtain
\begin{equation}
\phi_1^{{\rm TK2}^{AA}}(x)=\pm \frac{1}{2} \frac{\sinh 2 \sqrt{2}
(x+a)}{\cosh 2 \sqrt{2} (x+a)+b} \hspace{1cm} \phi^{{\rm
TK2}^{AA}}_2(x)=\pm \frac{1}{\sqrt{1+b^{-1} \cosh 2 \sqrt{2}
(x+a)}} \label{eq:sol12}
\end{equation}
where we have introduced $b=\frac{1}{\sqrt{1-4 c}}$. In the $b \in
(0,\infty)$ range, the above solutions are kinks that connect the
$A_1$ and $A_2$ vacua (see Figure 4). If $\sigma=\frac{1}{2}$,
(\ref{eq:traba1}) becomes $\phi_1^2+\frac{1}{2}
\phi_2^2=\frac{1}{4}+c \phi_2^{4}$, which  can be written as $(1+2
\phi_1-\phi_2^2)(1-2\phi_1-\phi_2^2)=0$ for $c=C^S=\frac{1}{4}$.
There are kinks on parabolic trajectories joining points in the
{\it A} and {\it B} vacuum orbits
\[
\phi_1^{{\rm TK2}^{AB}}(x)=\pm \frac{1}{4} \left(1-\tanh \sqrt{2}
(x+a) \right) \hspace{1cm} \phi_2^{{\rm TK2}^{AB}}(x)=\pm
 \sqrt{\frac{1}{
 2}\left(1+\tanh \sqrt{2} (x+a) \right)}
\]
and, replacing $x+a$ by $-x-a$, we obtain their antikinks.
\begin{figure}[htbp]
\centerline{\epsfig{file=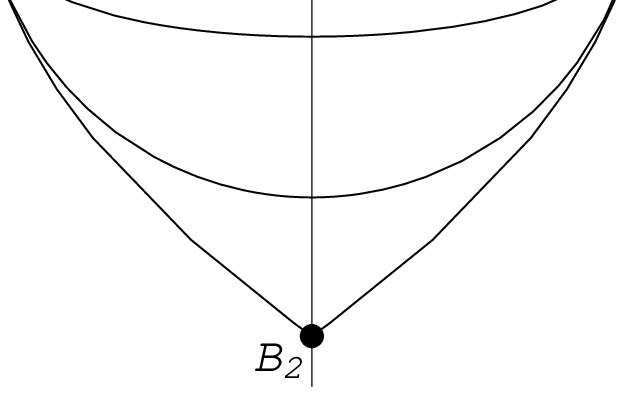, height=4.5cm} \hspace{0.5cm}
\epsfig{file=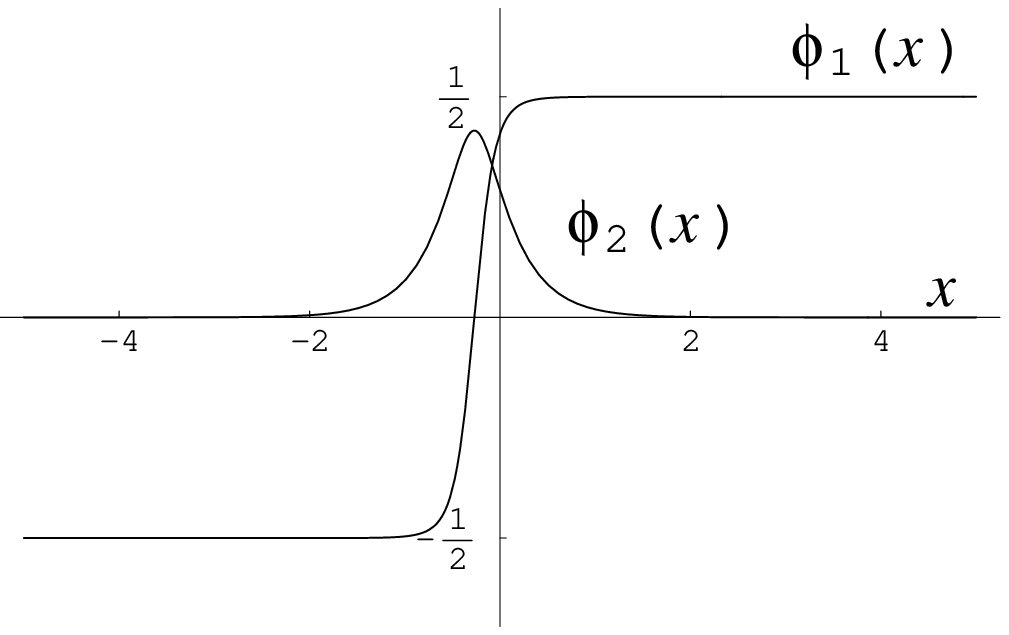,height=3.5cm}
\hspace{0.5cm}\epsfig{file=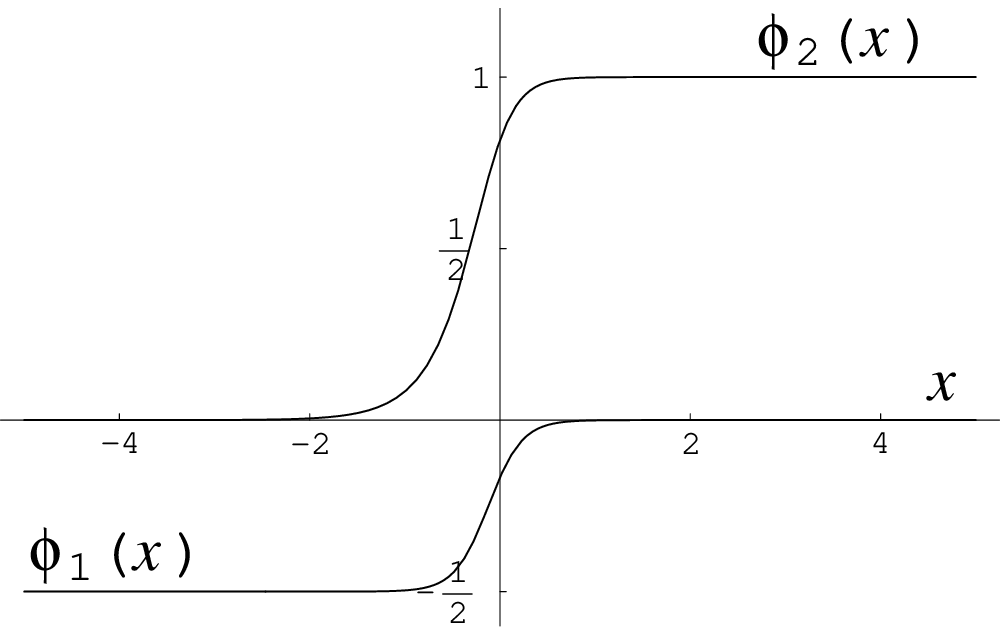, height=3.5cm}}
\caption{\small \it Kink curves (left), a kink in the AA sector
(middle) and a kink in the AB sector (right).}
\end{figure}

\section{The second superpotential: $\sigma=2$}

For $\sigma=2$, $U(\phi)=(4 \phi_1^2+4 \phi_2^2-1)^2+64 \phi_1^2
\phi_2^2$ does not change if we swap the field components. There
is a second superpotential in the model for $\sigma=2$:
$W'(\phi_1,\phi_2)=W(\phi_2,\phi_1)$. A second arrangement \'a la
Bogomol'nyi using $W'(\phi_1,\phi_2)$ provides another system of
first-order differential equations:
\[
\frac{d \phi_1}{d x}= \frac{\partial W'}{\partial\phi_1}=8
\sqrt{2} \phi_1 \phi_2 \hspace{1cm}\frac{d \phi_2}{d
x}=\frac{\partial W'}{\partial\phi_2}=\sqrt{2} (4 \phi_1^2+4
\phi_2^2-1)
\]
The flow-lines of ${\rm grad}\,W'$ connect $B_1$ and $B_2$, which
are respectively the maximum and the minimum of $W'$, whereas
$A_1$ and $A_2$ are $W'$ saddle points. We thus obtain a new
family of topological kinks, now in the {\it BB} sector, with the
r$\hat{{\rm o}}$les of $\phi_1$ and $\phi_2$ interchanged: if $b
\in (1,\infty)$,
\[
\phi_1^{{\rm TK2}^{BB}}(x)=\pm \frac{1}{2}
\frac{\sqrt{b^2-1}}{\cosh 4 \sqrt{2} (x+a)+b} \hspace{1cm}
\phi_2^{{\rm TK2}^{BB}}(x)=\pm \frac{1}{2} \frac{\sinh 4
\sqrt{2}(x+a)}{\cosh 4 \sqrt{2} (x+a)+b}
\]
are the two-component topological kinks in the {\it BB} sector. If
$c \rightarrow -\infty$ ($b \rightarrow 1$), we find the ${\rm
TK}1^{BB}$ kink and if $c=4$ ($b\rightarrow \infty$) the
separatrix kinks in the {\it AB} sector are reached at the
boundary of the component of the moduli space of kinks that
belong to the {\it BB} sector. The kink energy sum rules are:
$E_{{\rm TK}2^{AA}}= E_{{\rm TK}2^{BB}}=2E_{{\rm
TK}2^{AB}}=\frac{4}{3}a^3\lambda$.

\section{The moduli space of non-BPS kinks in the $BB$ topological
sector: $\sigma=\frac{1}{2}$}

If $\sigma=\frac{1}{2}$, there is also a second superpotential,
\begin{equation}
W'(\phi_1,\phi_2)=\frac{\sqrt{2}}{3} \sqrt{\phi_1^2+\phi_2^2}\,
\left(4 \phi_1^2+ \phi_2^2-3 \right)\quad , \label{eq:superpot12}
\end{equation}
that also solves the first equation in (\ref{eq:superpot}). The
second system of first-order equations
\begin{equation}
\frac{d \phi_1}{dx}=\pm\frac{\partial
W'}{\partial\phi_1}=\pm\frac{\sqrt{2} \phi_1 (4 \phi_1^2+3
\phi_2^2-1)}{\sqrt{\phi_1^2+\phi_2^2}} \hspace{1cm} \frac{d
\phi_2}{dx}=\pm\frac{\partial
W'}{\partial\phi_2}=\pm\frac{\sqrt{2} \phi_2 (2 \phi_1^2+
\phi_2^2-1)}{\sqrt{\phi_1^2+\phi_2^2}} \label{eq:ecudif12}
\end{equation}
rules the flows generated by $\pm{\rm grad}\,W'$ in the system.
$W'$ is not differentiable at the origin and the flows of $\pm{\rm
grad}\,W'$
\begin{equation}
\frac{d\phi_2}{d \phi_1}=\frac{\phi_2 (2 \phi_1^2+
\phi_2^2-1)}{\phi_1(4\phi_1^2+3\phi_2^2+1)}\label{eq:flow}
\end{equation}
are undefined at $O\equiv (0,0)\in{\Bbb R}^2$. Note that $B_1$ and
$B_2$ are both minima of $W'$, whereas $A_1$ and $A_2$ are $W'$
saddle points. The origin is the maximum of $W'$ and thus the
flow-lines of ${\rm grad}\,W'$ run from $O$ to either $B_1$ or
$B_2$. To obtain a kink orbit, we must glue at $O$ a $\gamma_-$
flow-line of ${\rm grad}\,W'$ with a $\gamma_+$ flow-line of
$-{\rm grad}\,W'$ smoothly. Because the flows are undefined at
$O$, we expect that an infinite number of lines will meet at the
origin.

The Bogomol\'ny splitting must to take this into account and the
energy of the kink solutions of (\ref{eq:ecudif12})
\begin{eqnarray*}
{\cal E}[\phi]&=& \int_{-a}^\infty  dx \, \frac{1}{2} \left\|
\frac{d \phi}{dx}- \frac{\partial W'}{\partial \phi} \right\|^2 +
\int_{-\infty}^{-a}  dx \, \frac{1}{2} \left\| \frac{d \phi}{dx}-
\frac{\partial W'}{\partial \phi} \right\|^2 +
T(\gamma_+)+T(\gamma_-)\\T&=&T(\gamma_+)+T(\gamma_-)=|W'(B_1)-W'(O)|+|W'(B_2)-W'(O)|
\end{eqnarray*}
${\cal E}[\phi_{{\rm TK}2^{BB}}]=T(\gamma_+)+T(\gamma_-)$ is not
topological; it depends on the value of the superpotential at the
origin, a sign of instability \cite{J0,J1}. The kink energy sum
rules are: $E_{{\rm TK}2^{BB}}=2\,E_{{\rm TK}2^{AA}}=4\,E_{{\rm
TK}2^{AB}}=\frac{8}{3}a^3\lambda$ and the ${\rm TK}2^{BB}$ kinks
decay to two  ${\rm TK}2^{AB}$ plus one  ${\rm TK}2^{AA}$ kinks.

Using parabolic variables, we have shown  that the integration of
(\ref{eq:ecudif12}) reduces to quadratures in Reference
\cite{Aai2}. The translation of our results to Cartesian
coordinates is as follows:
\begin{itemize}

\item The kink orbits that solve (\ref{eq:flow}) satisfy the
equation
\begin{equation}
16 \, e^{4 \sqrt{2} c} \, \phi_1^2 \, (\phi_1^2+\phi_2^2)+(1-e^{4
\sqrt{2} c})^2 \, \phi_2^4 \, (2 \phi_1-\phi_2^2+1)(2
\phi_1+\phi_2^2-1)=0 \label{eq:orb12}
\end{equation}
and are plotted in Figure 5. Here $c$ is a real integration
constant. \noindent\begin{figure}[htbp]
\centerline{\epsfig{file=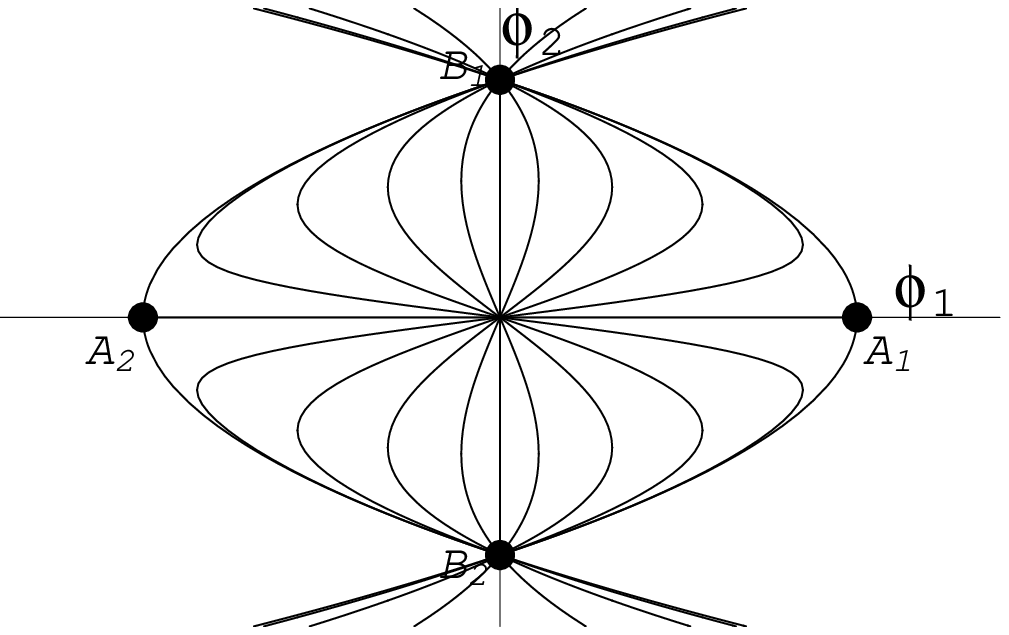, height=4.cm} \hspace{1cm}
\epsfig{file=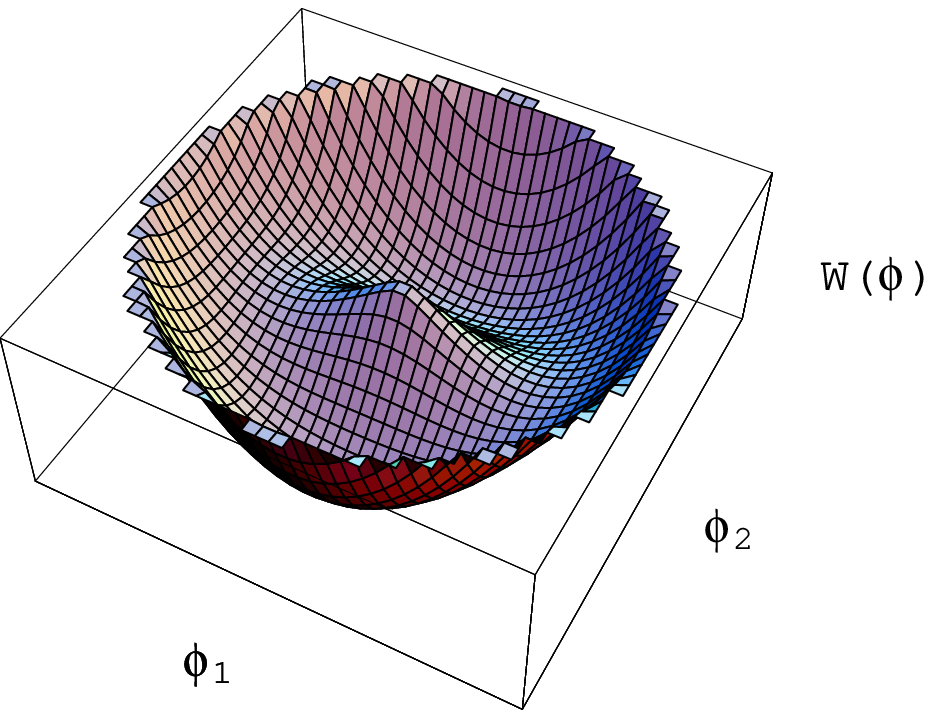, height=4.cm} } \caption{\small
TK2$^{BB}(c)$ \it Kink family (left) and the superpotential $W'$
(right) }
\end{figure}
\item Analytically, the variety of ${\rm TK}2^{BB}(c)$ kinks is
given by:
\begin{equation}
\phi_1^{{\rm TK}2^{BB}}(x)= \frac{\sinh 2 \sqrt{2}c \, \sinh 2
\sqrt{2}(x+a)}{\cosh^2 2 \sqrt{2}(x+a) + 2 \cosh 2 \sqrt{2} c\cosh
2 \sqrt{2}(x+a) +1} \label{eq:tk212a}
\end{equation}
\begin{equation}
\phi_2^{{\rm TK}2^{BB}}(x)= \frac{-\sinh 2
\sqrt{2}(x+a)}{\sqrt{\cosh^2 2 \sqrt{2} (x+a) + 2 \cosh 2 \sqrt{2}
c \cosh 2 \sqrt{2}(x+a) +1}} \label{eq:tk212b}
\end{equation}
Besides the soliton center $x=-a$, the kink family is parametrized
by c.
\end{itemize}
Because the spatial translations $T_a:x\to x+a$ leads from one
solution to another and
\[
\pi_1(\phi_1^{{\rm TK}2^{BB}}(x;c),\phi_2^{{\rm
TK}2^{BB}}(x;c))=(\phi_1^{{\rm TK}2^{BB}}(x;-c),\phi_2^{{\rm
TK}2^{BB}}(x;-c)) \qquad ,
\]
the moduli space of ${\rm TK}2^{BB}$ kinks -the quotient of the
(\ref{eq:tk212a}-\ref{eq:tk212b}) kink variety by the action of
$T_a$ and $\pi_1$- is the open half-line: $c\in (0,\infty)$. If,
moreover, we take quotient by $P:x+a\to -x-a$, the anti-kinks are
also included in the moduli space.

The asymptotic behaviour
\[
\lim_{x\rightarrow\pm\infty}\phi_1^{{\rm TK}2^{BB}}(x;c)=0 \qquad
, \qquad \lim_{x\rightarrow\pm\infty}\phi_2^{{\rm
TK}2^{BB}}(x;c)=\mp 1
\]
fits in with the boundary behaviour, guaranteeing finite energy to
the ${\rm TK}2^{BB}(c)$ kinks. They are not stable because all of
them cross the origin:
\[
\phi_1^{{\rm TK}2^{BB}}(-a;c)=0 \qquad , \qquad \phi_2^{{\rm
TK}2^{BB}}(-a;c)=0 \quad .
\]
Thus, only if $x\in (-\infty, -a)$
(\ref{eq:tk212a}-\ref{eq:tk212b}) are solutions of the
first-order equations (\ref{eq:ecudif12}) with the + sign, whereas
they solve (\ref{eq:ecudif12}) with the - sign in the $x\in
(-a,\infty)$ range, or vice-versa. It can easily be proved,
however, that these solutions satisfy the second-order
differential equations (\ref{eq:ecuseg}).

Things are different at the boundary of the moduli space, the
union of the $c=0$ and $c=\infty$ points. Looking at the formula
(\ref{eq:orb12}) we find the ${\rm TK}1^{BB}$ kink as the $c=0$
limit of the kink variety, whereas the ${\rm TK}1^{AA}$ kink and
two ${\rm TK}^{AB}$ kinks -that live on different parabolic
branches- are met at $c=\infty$.

\end{document}